\documentclass[onecolumn]{aastex62}
\pdfoutput=1 
\usepackage{amsmath,amstext}
\usepackage[T1]{fontenc}
\usepackage{apjfonts} 
\usepackage[figure,figure*]{hypcap}
\usepackage[utf8]{inputenc}
\usepackage{array}
\usepackage{makecell}
\usepackage{mathrsfs}
\usepackage{mathtools}
\usepackage{footmisc}

\shorttitle{Search for Neutrino Emission from FRBs}
\shortauthors{M. Aartsen et al.}

\begin{document}
\email{analysis@icecube.wisc.edu}
\title{A Search for MeV to TeV Neutrinos from Fast Radio Bursts with IceCube}

\affiliation{III. Physikalisches Institut, RWTH Aachen University, D-52056 Aachen, Germany} \affiliation{Department of Physics, University of Adelaide, Adelaide, 5005, Australia} \affiliation{Dept. of Physics and Astronomy, University of Alaska Anchorage, 3211 Providence Dr., Anchorage, AK 99508, USA} \affiliation{Dept. of Physics, University of Texas at Arlington, 502 Yates St., Science Hall Rm 108, Box 19059, Arlington, TX 76019, USA} \affiliation{CTSPS, Clark-Atlanta University, Atlanta, GA 30314, USA} \affiliation{School of Physics and Center for Relativistic Astrophysics, Georgia Institute of Technology, Atlanta, GA 30332, USA} \affiliation{Dept. of Physics, Southern University, Baton Rouge, LA 70813, USA} \affiliation{Dept. of Physics, University of California, Berkeley, CA 94720, USA} \affiliation{Lawrence Berkeley National Laboratory, Berkeley, CA 94720, USA} \affiliation{Institut f{\"u}r Physik, Humboldt-Universit{\"a}t zu Berlin, D-12489 Berlin, Germany} \affiliation{Fakult{\"a}t f{\"u}r Physik {\&} Astronomie, Ruhr-Universit{\"a}t Bochum, D-44780 Bochum, Germany} \affiliation{Universit{\'e} Libre de Bruxelles, Science Faculty CP230, B-1050 Brussels, Belgium} \affiliation{Vrije Universiteit Brussel (VUB), Dienst ELEM, B-1050 Brussels, Belgium} \affiliation{Dept. of Physics, Massachusetts Institute of Technology, Cambridge, MA 02139, USA} \affiliation{Dept. of Physics and Institute for Global Prominent Research, Chiba University, Chiba 263-8522, Japan} \affiliation{Dept. of Physics and Astronomy, University of Canterbury, Private Bag 4800, Christchurch, New Zealand} \affiliation{Dept. of Physics, University of Maryland, College Park, MD 20742, USA} \affiliation{Dept. of Astronomy, Ohio State University, Columbus, OH 43210, USA} \affiliation{Dept. of Physics and Center for Cosmology and Astro-Particle Physics, Ohio State University, Columbus, OH 43210, USA} \affiliation{Niels Bohr Institute, University of Copenhagen, DK-2100 Copenhagen, Denmark} \affiliation{Dept. of Physics, TU Dortmund University, D-44221 Dortmund, Germany} \affiliation{Dept. of Physics and Astronomy, Michigan State University, East Lansing, MI 48824, USA} \affiliation{Dept. of Physics, University of Alberta, Edmonton, Alberta, Canada T6G 2E1} \affiliation{Erlangen Centre for Astroparticle Physics, Friedrich-Alexander-Universit{\"a}t Erlangen-N{\"u}rnberg, D-91058 Erlangen, Germany} \affiliation{Physik-department, Technische Universit{\"a}t M{\"u}nchen, D-85748 Garching, Germany} \affiliation{D{\'e}partement de physique nucl{\'e}aire et corpusculaire, Universit{\'e} de Gen{\`e}ve, CH-1211 Gen{\`e}ve, Switzerland} \affiliation{Dept. of Physics and Astronomy, University of Gent, B-9000 Gent, Belgium} \affiliation{Dept. of Physics and Astronomy, University of California, Irvine, CA 92697, USA} \affiliation{Karlsruhe Institute of Technology, Institut f{\"u}r Kernphysik, D-76021 Karlsruhe, Germany} \affiliation{Dept. of Physics and Astronomy, University of Kansas, Lawrence, KS 66045, USA} \affiliation{SNOLAB, 1039 Regional Road 24, Creighton Mine 9, Lively, ON, Canada P3Y 1N2} \affiliation{Department of Physics and Astronomy, UCLA, Los Angeles, CA 90095, USA} \affiliation{Department of Physics, Mercer University, Macon, GA 31207-0001} \affiliation{Dept. of Astronomy, University of Wisconsin, Madison, WI 53706, USA} \affiliation{Dept. of Physics and Wisconsin IceCube Particle Astrophysics Center, University of Wisconsin, Madison, WI 53706, USA} \affiliation{Institute of Physics, University of Mainz, Staudinger Weg 7, D-55099 Mainz, Germany} \affiliation{Department of Physics, Marquette University, Milwaukee, WI, 53201, USA} \affiliation{Institut f{\"u}r Kernphysik, Westf{\"a}lische Wilhelms-Universit{\"a}t M{\"u}nster, D-48149 M{\"u}nster, Germany} \affiliation{Bartol Research Institute and Dept. of Physics and Astronomy, University of Delaware, Newark, DE 19716, USA} \affiliation{Dept. of Physics, Yale University, New Haven, CT 06520, USA} \affiliation{Dept. of Physics, University of Oxford, Parks Road, Oxford OX1 3PU, UK} \affiliation{Dept. of Physics, Drexel University, 3141 Chestnut Street, Philadelphia, PA 19104, USA} \affiliation{Physics Department, South Dakota School of Mines and Technology, Rapid City, SD 57701, USA} \affiliation{Dept. of Physics, University of Wisconsin, River Falls, WI 54022, USA} \affiliation{Dept. of Physics and Astronomy, University of Rochester, Rochester, NY 14627, USA} \affiliation{Oskar Klein Centre and Dept. of Physics, Stockholm University, SE-10691 Stockholm, Sweden} \affiliation{Dept. of Physics and Astronomy, Stony Brook University, Stony Brook, NY 11794-3800, USA} \affiliation{Dept. of Physics, Sungkyunkwan University, Suwon 16419, Korea} \affiliation{Dept. of Physics and Astronomy, University of Alabama, Tuscaloosa, AL 35487, USA} \affiliation{Dept. of Astronomy and Astrophysics, Pennsylvania State University, University Park, PA 16802, USA} \affiliation{Dept. of Physics, Pennsylvania State University, University Park, PA 16802, USA} \affiliation{Dept. of Physics and Astronomy, Uppsala University, Box 516, S-75120 Uppsala, Sweden} \affiliation{Dept. of Physics, University of Wuppertal, D-42119 Wuppertal, Germany} \affiliation{DESY, D-15738 Zeuthen, Germany}  \author{M. G. Aartsen} \affiliation{Dept. of Physics and Astronomy, University of Canterbury, Private Bag 4800, Christchurch, New Zealand} \author{M. Ackermann} \affiliation{DESY, D-15738 Zeuthen, Germany} \author{J. Adams} \affiliation{Dept. of Physics and Astronomy, University of Canterbury, Private Bag 4800, Christchurch, New Zealand} \author{J. A. Aguilar} \affiliation{Universit{\'e} Libre de Bruxelles, Science Faculty CP230, B-1050 Brussels, Belgium} \author{M. Ahlers} \affiliation{Niels Bohr Institute, University of Copenhagen, DK-2100 Copenhagen, Denmark} \author{M. Ahrens} \affiliation{Oskar Klein Centre and Dept. of Physics, Stockholm University, SE-10691 Stockholm, Sweden} \author{C. Alispach} \affiliation{D{\'e}partement de physique nucl{\'e}aire et corpusculaire, Universit{\'e} de Gen{\`e}ve, CH-1211 Gen{\`e}ve, Switzerland} \author{K. Andeen} \affiliation{Department of Physics, Marquette University, Milwaukee, WI, 53201, USA} \author{T. Anderson} \affiliation{Dept. of Physics, Pennsylvania State University, University Park, PA 16802, USA} \author{I. Ansseau} \affiliation{Universit{\'e} Libre de Bruxelles, Science Faculty CP230, B-1050 Brussels, Belgium}
% ...
\date{\today}  \collaboration{IceCube Collaboration} \noaffiliation

\begin{abstract}
We present two searches for IceCube neutrino events coincident with 28 fast radio bursts (FRBs) and one repeating FRB. The first improves upon a previous IceCube analysis~\textendash~searching for spatial and temporal correlation of events with FRBs at energies greater than roughly 50 GeV~\textendash~by increasing the effective area by an order of magnitude. The second is a search for temporal correlation of MeV neutrino events with FRBs. No significant correlation is found in either search, therefore, we set upper limits on the time-integrated neutrino flux emitted by FRBs for a range of emission timescales less than one day. These are the first limits on FRB neutrino emission at the MeV scale, and the limits set at higher energies are an order-of-magnitude improvement over those set by any neutrino telescope.\\
\end{abstract}

\section{Introduction}
The IceCube Neutrino Observatory instruments 1 km$^3$ of Antarctic ice between depths of 1450 and 2450 meters at the geographic South Pole and records particle interactions in the ice by capturing Cherenkov radiation, produced by secondary particles,  in photomultiplier tubes (PMT) that are housed in a glass pressure vessel known as digital optical modules (DOMs)~\citep{Aartsen:2016nxy}. IceCube observes a cosmic neutrino flux from 10 TeV to a few PeV \citep{Aartsen:2015knd, Aartsen:2015rwa} and  recently reported evidence for neutrino emission from the blazar TXS 0506+056 \citep{IceCube:2018dnn, IceCube:2018cha}. 

Despite the evidence for neutrino emission from TXS 0506+056, the overwhelming majority of the diffuse astrophysical flux remains unexplained. Transient sources play a major role in high-energy astrophysics and potentially could account for a large fraction of the detected neutrino flux. Fast radio bursts (FRBs) are a class of non-periodic, highly dispersed, millisecond-scale radio flashes \citep{Lorimer:2018rwi, Keane:2018jqo}. Although fewer than 60 unique sources have been detected, the expected rate of detectable FRBs each day is in the thousands \citep{Bhandari:2017qrj}. Many models for FRBs have been offered \citep{Platts:2018hiy}, but due to the scarcity of discoveries and multi-wavelength follow-up detections, none is strongly favored. Some of these models allow for hadronic acceleration in the vicinity of the progenitors, such as super massive neutron stars,  or supernova explosions, which would lead to the production of both high-energy cosmic rays and neutrinos \citep{Li:2013kwa, DasGupta:2017uac}. 

Previous analyses have set upper limits on neutrino emission from fast radio bursts using quality track-like events in IceCube \citep{Fahey:2016czk, FRB_6yr}. A search for multiplets of track-like neutrino events with minute-scale temporal coincidence limits the number density of a transient source class, under particular source evolution assumption, producing IceCube astrophysical flux to larger than $10^{-5}\mathrm{~Mpc}^{-3}\mathrm{yr}^{-1}$ \citep{Aartsen:2018fpd}; the numerous and dim emission from FRBs is consistent with this lower bound \citep{Callister:2016vtl}.

This work includes two analyses that improve existing constraints on neutrino emission from FRBs and test a wider range of neutrino energies. IceCube has access to two different energy ranges and we take advantage of that to search for coincidences with FRBs. In Section~\ref{sec:tracks}, we present a search using a track-like event selection with improved effective area compared to the previous IceCube searches. In Section~\ref{sec:sndaq}, we present a search for temporal correlation of MeV neutrinos with FRBs. Section~\ref{sec:conclusion} summarizes analysis results and discusses the outlook for future searches for neutrinos from FRBs.\\

\section{Search for Coincident Muon Track Events}
\label{sec:tracks}

This search uses a procedure similar to that of IceCube's previous search for neutrino emission from FRBs \citep{FRB_6yr}, that is hereafter referred to as the \emph{six-year analysis}. These analyses search for both temporal and spatial correlation of FRBs and muon neutrino events, in which a muon created from a charged-current interaction leaves a track-like signature in the detector. All 28 non-repeating FRBs are analyzed (Table~\ref{tab:FRBs}) in a source-stacking search and in a search for the brightest source. The six-year analysis used an event selection that was initially optimized for analyses of gamma-ray bursts. Because FRBs have much shorter durations, higher levels of background are tolerated and we can use a looser event selection procedure to increase acceptance to astrophysical muon neutrinos and improve analysis sensitivity at emission timescales less than 10$^3$ seconds.

\subsection{Event sample}
\label{sec:event_selection}

The data used in this analysis consist of through-going muon neutrino candidate events from 2011-02-18 through 2018-03-13. In the six-year analysis, in order to reduce the fraction of atmospheric muons in the sample, the data consisted largely of events with very high energy (${E_{\nu}>10\mathrm{~TeV}}$) or that had penetrated many kilometers of ice prior to detection. Here, we instead use an event selection closer to IceCube's trigger level, resulting in a higher rate of atmospheric muons but increasing the acceptance of astrophysical neutrinos as well. This event selection focuses on removing low-energy events so that passing data can be transmitted via satellite within the bandwidth limit of roughly 75 gigabytes per day.

The sample of muon track events has an average all-sky rate of 35.7 Hz due mainly to penetrating muons from cosmic ray interactions in Earth's atmosphere. Roughly 5 events per day are caused by astrophysical muon neutrinos\footnote{This estimate is calculated by combining the effective area of the event selection with IceCube's global fit of the diffuse astrophysical neutrino flux \citep{Aartsen:2015knd}, which is an $E^{-2.49}$ unbroken power-law.}.

\begin{figure}
\centering
  \begin{minipage}{0.48\textwidth}
  \centering
  \includegraphics[width=\textwidth]{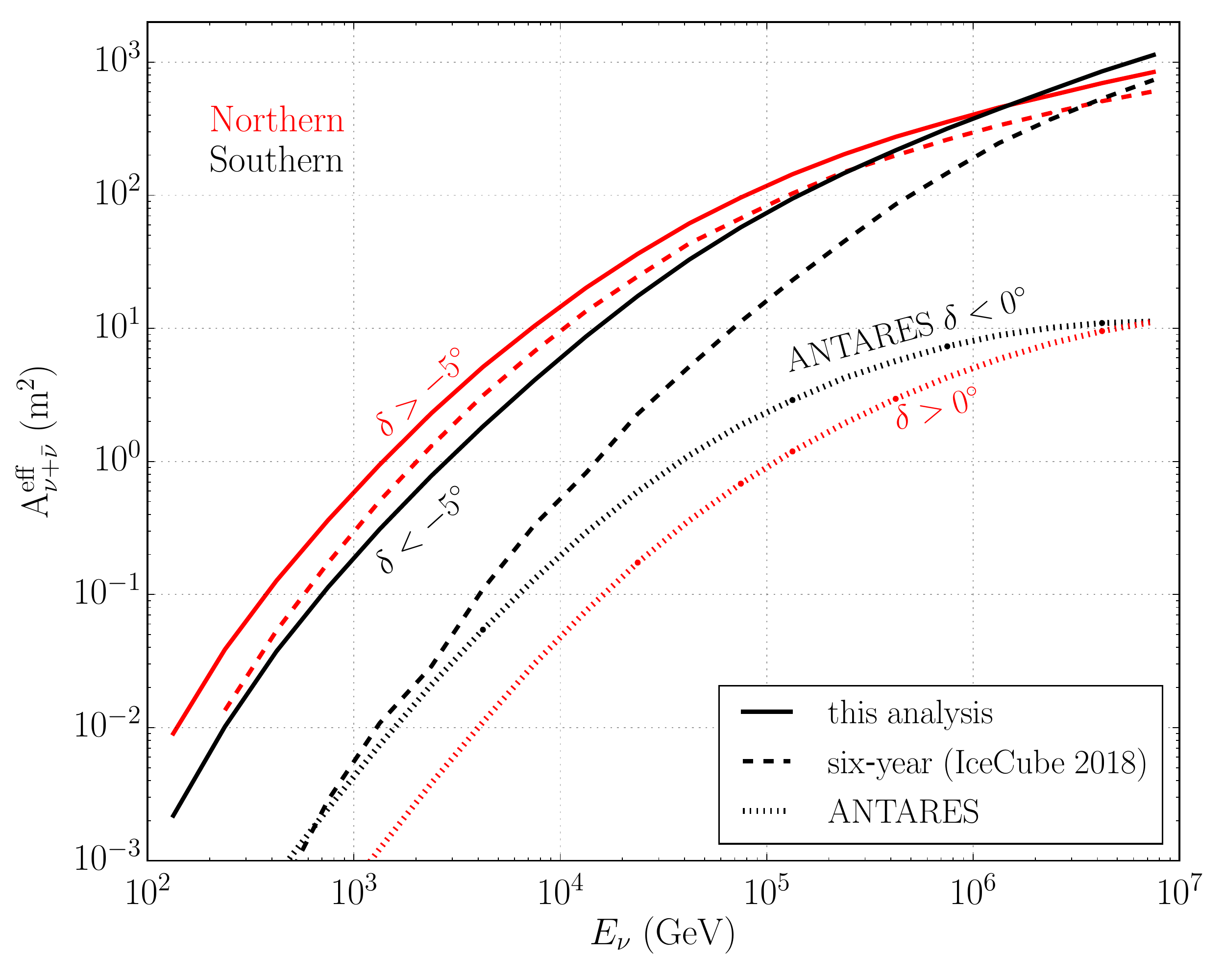}
  \caption{With an event selection looser than in IceCube's six-year search for neutrinos from FRBs \citep{FRB_6yr}, the effective area of this selection to muon neutrinos is improved significantly, especially in the Southern sky (black) and at energies less than 1 PeV. This is compared to the smaller effective area of ANTARES for a point-source event selection \citep{ANTARES1} that approximately reproduces the effective area from their 2018 FRB analysis (Figure 5 in \citep{ANTARES2}).}
  \end{minipage}
  \label{fig:eff_area}
  \hfill
  \begin{minipage}{0.48\textwidth}  
  \centering 
  \includegraphics[width=\textwidth]{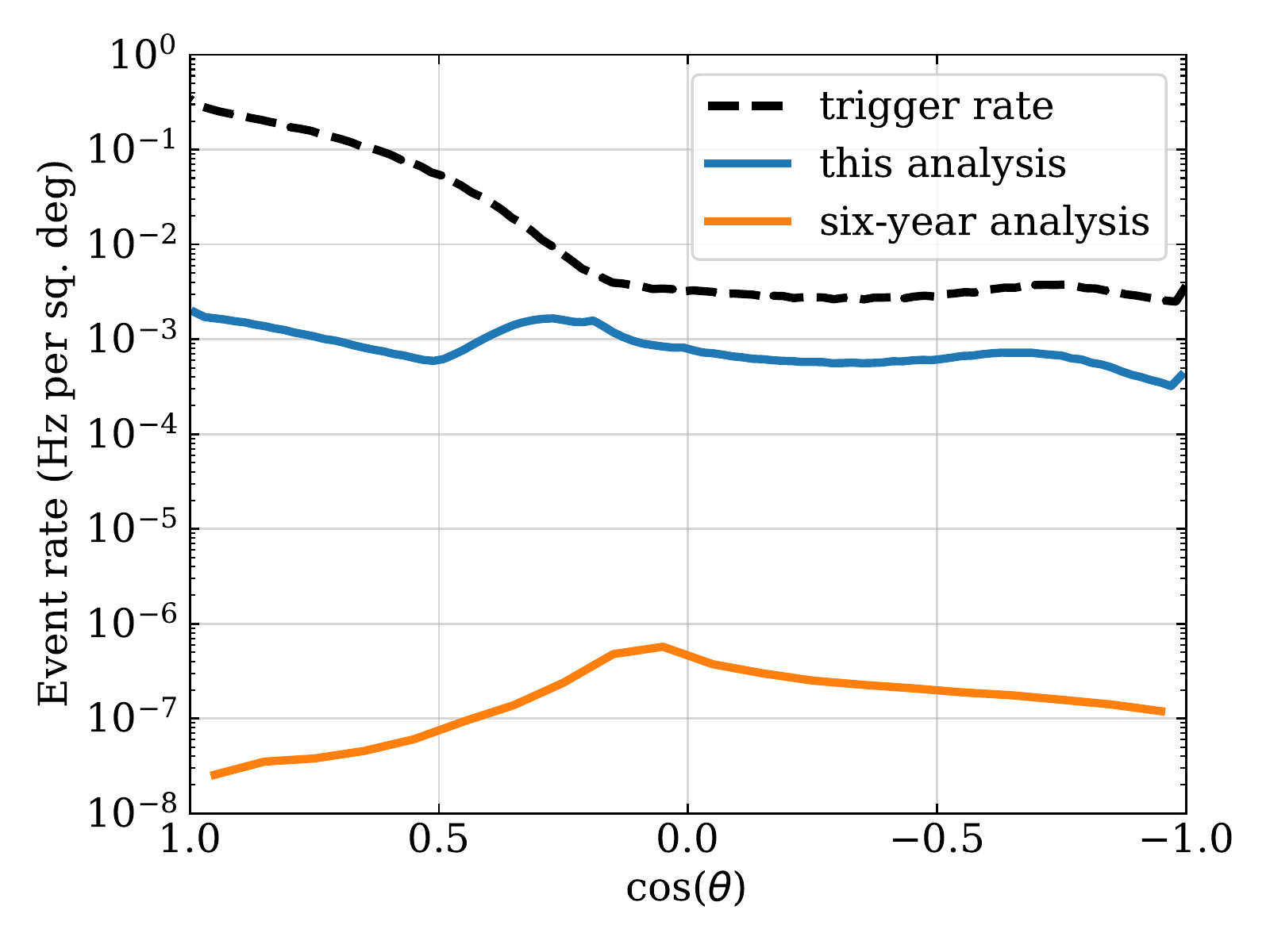}
  \caption{The event rate for this data is about 4000$\times$ larger than that of the six-year analysis data for almost all zenith angles, $\theta$, due mostly to an increase in the number of muons generated by cosmic rays passing the event selection. The increase is largest in the Southern sky (${\cos(\theta)>0}$), approaching a factor of $10^5$ increase at zenith. Features in the Southern sky are the result of selection methods; e.g., the deposited energy required to pass event selection is higher for events with ${\theta < 70^{\circ}}$.}
  \end{minipage}
  \label{fig:zenith_pdf}
  \vspace{0.2in}
\end{figure}

Compared to the six-year analysis, the effective area of this event selection to muon neutrinos is an order of magnitude larger in the Southern Sky (Figure~\ref{fig:eff_area}), with the largest improvements coming from energies below 100 TeV. In the Northern Sky, where the Earth already attenuates the atmospheric muon background, the average improvement is roughly 50\% in effective area. Correspondingly, the background rate for this event selection is much larger than for the six-year analysis. Figure~\ref{fig:zenith_pdf} compares the distribution of background data versus zenith angle for the six-year analysis to this event selection; the peak-to-peak seasonal variation in this rate is about 25\%, with a maximum in January, the austral summer\footnote{ Seasonal variation in atmospheric density affects the fraction of cosmic-ray-produced pions and kaons that decay, producing ice-penetrating muons, before otherwise interacting in the air. This causes IceCube's trigger rate due to atmospheric muons to peak in the austral summer. For more details on seasonal variation, see \citep{Grashorn:2009ey,2010arXiv1001.0776T, Desiati:2011hea} and references therein.}.

The total of 28 non-repeating FRBs were analyzed in this search. The information about each FRB is presented in Table~\ref{tab:FRBs}. 

\subsection{Analysis method}

The test statistic (TS) defined here is similar to that of the six-year analysis \citep{FRB_6yr}. For a search time window ${\Delta T}$, temporal correlation of an event with an FRB is satisfied if the event triggers the detector in the interval ${[t_{FRB}-\Delta T /2,~~ t_{FRB}+\Delta T /2]}$. Using a model-independent maximum likelihood method, events temporally coincident with a single FRB contribute to the maximum likelihood ratio test statistic, defined as
\begin{equation}
\textrm{TS}= -{\hat{n}_s}+\sum_{i=1}^{N}\ln\Big[1+\frac{{\hat{n}_s}S(x_i)}{n_b B(x_i)}\Big] \; ,
\label{TS_def}
\end{equation}

 where $S(x_i)$ is the total spatial probability density distribution (PDF) that considers the angular distance of an event direction $x_i$ with respect to the coordinates of a given FRB and $B(x_i)$ combines separate spatial and temporal parameterizations of data to describe the background PDF in that time and direction. 
 Here, $n_b$ is the expected number of background events in ${\Delta T}$, and likelihood is maximized with respect to the best fit number of observed signal events, $\hat{n}_s$. In the stacking search, the test statistic in Eq. \ref{TS_def} will have an additional sum over the number of FRBs in the search.

As in the six-year analysis, two tests are performed: The stacking test, which tests the hypothesis that the astrophysical class of FRBs emits neutrinos, evaluates the TS for all events in ${\Delta T}$ centered on all sources; the max-burst test, which tests the hypothesis that among a heterogeneous class of FRBs, one or a few bright sources emit neutrinos,  evaluates a TS separately for each FRB and its respective events, returning only the largest TS as the observation at ${\Delta T}$. In the stacking test, we consider a range of neutrino emission timescales by evaluating expanding iterations of ${\Delta T}$ from 0.03 s to 10$^5$ s: ${\Delta T = 0.01 \cdot 10^{i/2}\mathrm{~s,~where~}i = 1,2,...,14}$. Beyond ${\Delta T > 10^3 \mathrm{~s}}$, the sensitivity of the max-burst analysis exceeds upper limits set in the six-year analysis for all tested spectra, therefore we do not evaluate larger ${\Delta T}$ for the max-burst test.

\subsection{Results}

We find the results from both the stacking test and max-burst test consistent with the background-only hypothesis (Table~\ref{tab:Level2_TS_results}). After trials-correcting each test for the number of time windows searched\footnote{Test results from consecutive time windows are correlated in this analysis, as the smaller time windows are contained within the larger time windows. Therefore, the trials factor is less than the number of windows searched. A Monte Carlo simulation calculates the probability of exceeding the smallest pre-trial $p$-value over the course of expansion of $\Delta T$ -- this probability is the post-trial $p$-value.}, the $p$-values for the stacking and max-burst tests are 0.35 and 0.33, respectively.

\begin{table}
	\caption{ We show the test statistic (TS) values for the stacking and max-burst tests in each time window of the high-energy (Muon Tracks) search. The median TS from 10$^9$ trials of background-only simulation are shown for comparison, along with pre-trial $p$-values for results in their respective time windows. For the max-burst analysis, we only test the $\Delta T$ in which there is an improvement in sensitivity relative to the six-year analysis. Both tests produce most significant results in ${\Delta T = 10^3 \mathrm{~s}}$; post-trials significance for the stacking and max-source tests are ${p=0.35}$ and ${p=0.33}$ respectively.}
	\centering
	\begin{tabular}{ c|c c c|c c c}
	    \hline
	    \hline
		& & Stacking TS & & &  Max-source TS & \\
		$\Delta T$ (s) & median & \textbf{result} & $p$ & median & \textbf{result} & $p$ \\
		\hline
		\hline
		3.16e-2 & 0 & 0 & 1 & 1.13 & 0.32 & 0.87 \\
		\hline
		1.00e-1 & 0 & 0 & 1 & 1.21 & 1.10 & 0.54 \\
		\hline
		3.16e-1 & 0 & 0 & 1 & 1.30 & 1.31 & 0.49 \\
		\hline
		1.00e-0 & 0 & 0 & 1 & 1.40 & 0.12 & 0.98 \\
		\hline
		3.16e-0 & 0 & 0 & 1 & 1.54 & 1.35 & 0.57 \\
		\hline
		1.00e+1 & 0 & 0.29 & 0.124 & 1.76 & 2.38 & 0.32 \\
		\hline
		3.16e+1 & 0 & 0.02 & 0.256 & 2.03 & 2.85 & 0.27 \\
		\hline
		1.00e+2 & 0 & 0.07 & 0.274 & 2.44 & 4.86 & 0.07 \\
		\hline
		3.16e+2 & 0 & 0 & 1 & 3.22 & 6.06 & 0.08 \\
		\hline
		1.00e+3 & 0.024 & 2.32 & 0.042 & 5.05 & 10.57 & 0.05 \\
		\hline
		3.16e+3 & 0.208 & 1.64 & 0.141 & - & - & - \\
		\hline
		1.00e+4 & 0.779 & 0.79 & 0.492 & - & - & - \\
		\hline
		3.16e+4 & 2.559 & 0 & 1 & - & - & - \\
		\hline
		1.00e+5 & 8.023 & 0 & 1 & - & - & - \\
		\hline
	\end{tabular}
	\label{tab:Level2_TS_results}
	\vspace{0.2in}
\end{table}

\begin{figure*}
\centering
  \begin{minipage}[b]{0.48\textwidth}
  \centering
  \includegraphics[width=\textwidth]{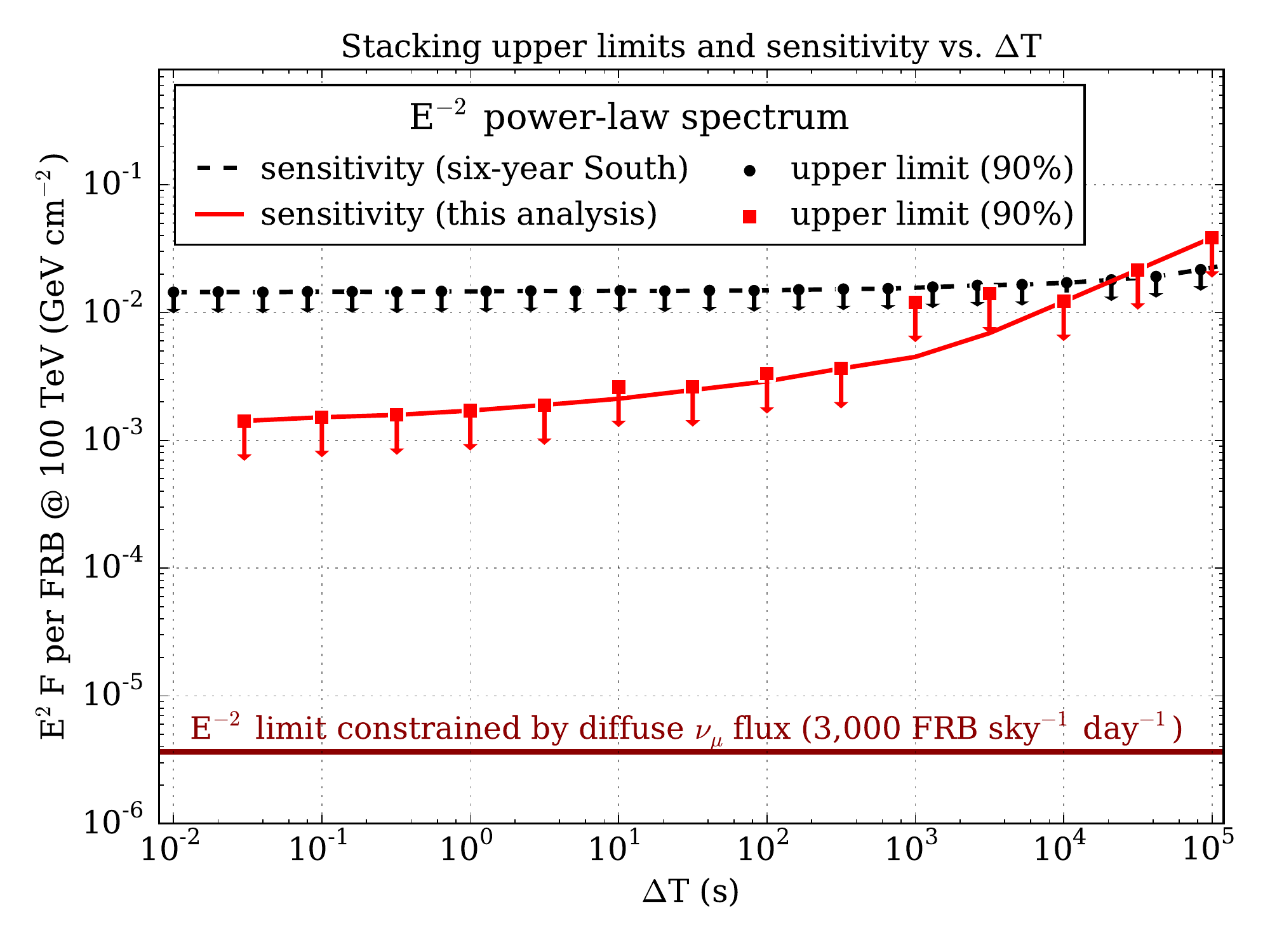}
  \label{fig:stack_E2}
  \end{minipage}
  \quad
  \begin{minipage}[b]{0.48\textwidth}  
  \centering 
  \includegraphics[width=\textwidth]{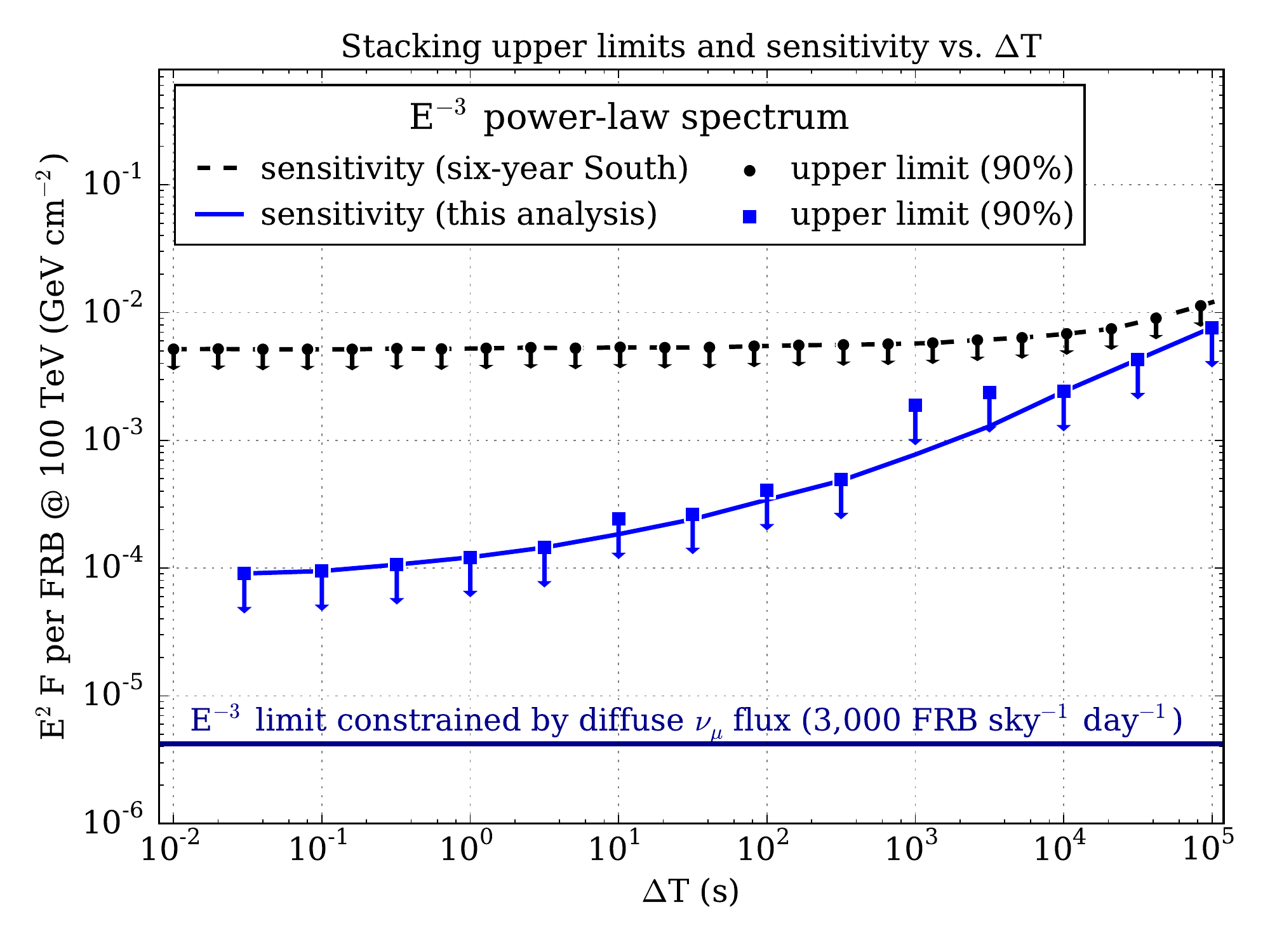}
  \label{fig:stack_E3}
  \end{minipage}
  \quad
  \begin{minipage}[b]{0.48\textwidth}   
  \centering 
  \includegraphics[width=\textwidth]{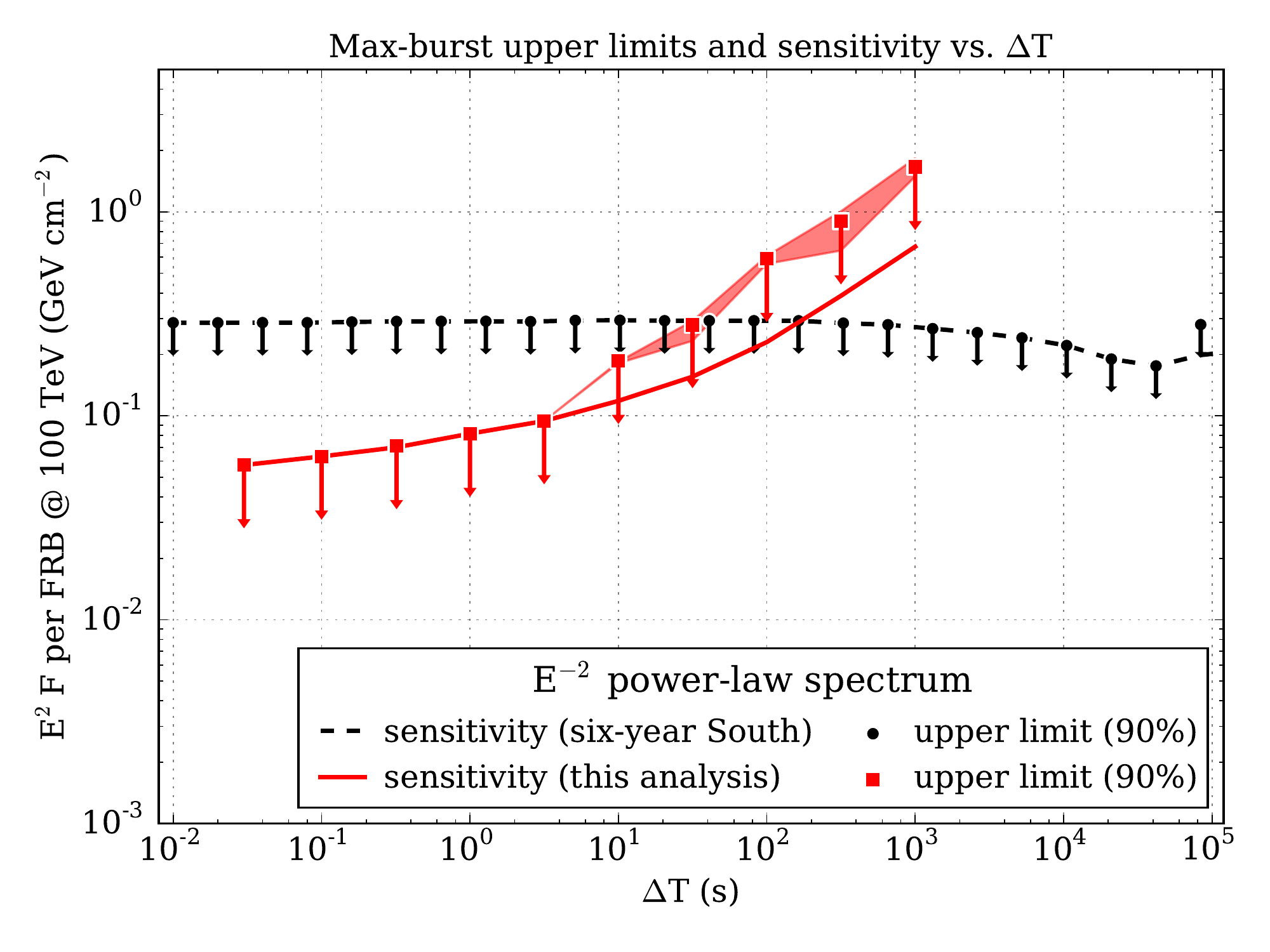}
  \label{fig:max_E2}
  \end{minipage}
  \quad
  \begin{minipage}[b]{0.48\textwidth}   
  \centering 
  \includegraphics[width=\textwidth]{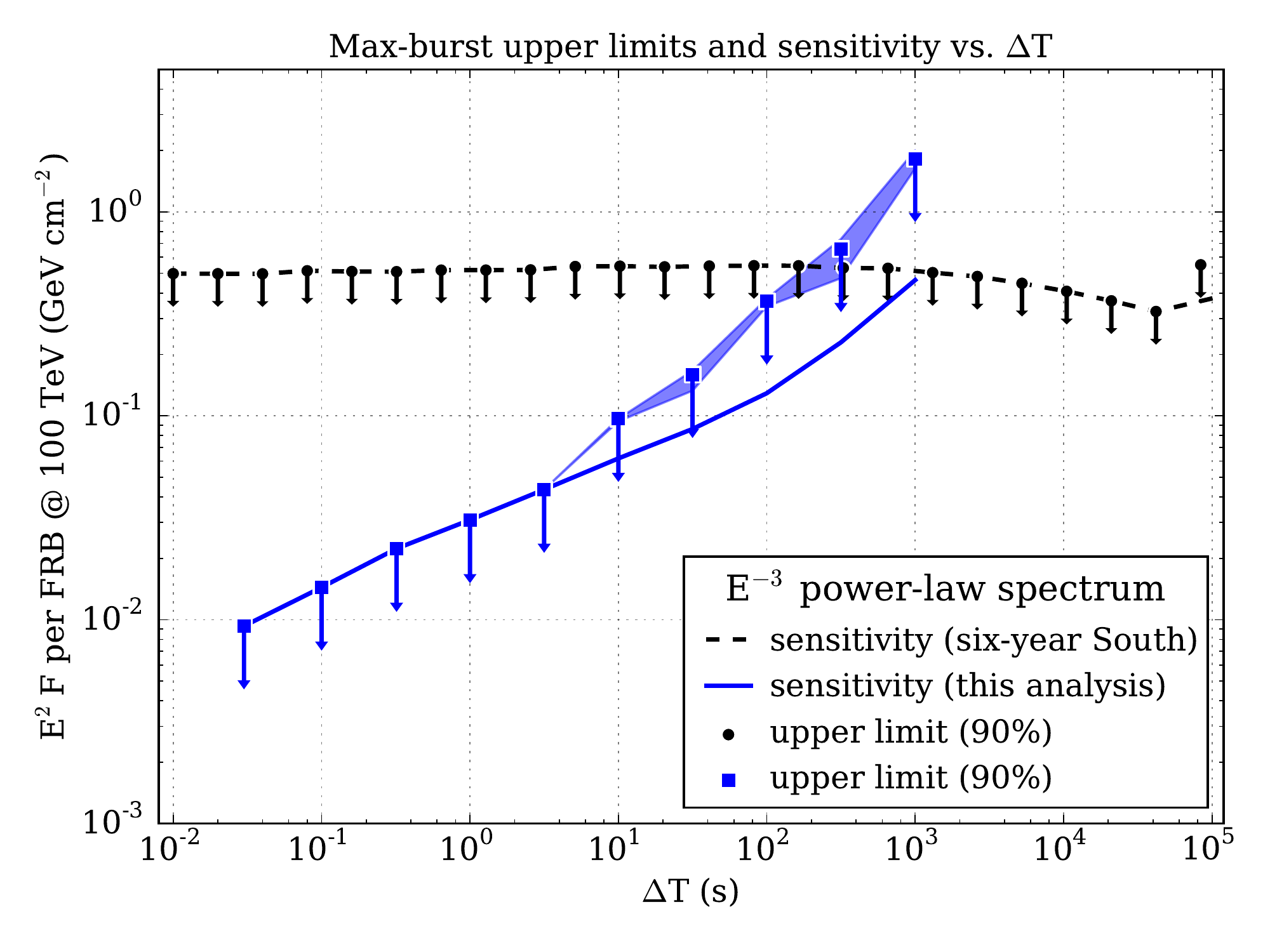}
  \label{fig:max_E3}
  \end{minipage}
  \caption{We set upper limits on the time-integrated neutrino flux per FRB for a range of $\Delta T$, assuming power-law spectra of $E^{-2}$ (top left) and $E^{-3}$ (top right). These limits provide an order-of-magnitude improvement over the previous best limits on non-repeating FRBs \citep{FRB_6yr}. For comparison, we show constraints produced by dividing IceCube’s entire astrophysical $\nu_{\mu}$ flux \citep{Aartsen:2017mau} equally among a homogeneous class of 3,000 FRBs per day. We also set upper limits on the maximum time-integrated neutrino flux among 28 FRBs for every $\Delta T$, assuming power-law spectra of $E^{-2}$ (bottom left) and $E^{-3}$ (bottom right). The error bands on these limits represents the central 90\% of systematic variation in limits due to uncertainty in background parameterization.}
  \label{fig:limits}
\end{figure*}

Upper limits are calculated (90\% confidence level) for the time-integrated flux per FRB at each $\Delta T$ (Figure~\ref{fig:limits}). The results are listed in Table~\ref{tab:limits_perFRB}. In the stacking search (top panels), the limits we set for ${\Delta T < 1 \mathrm{~s}}$ are factors of 10 and 50 stronger on spectra of $E^{-2}$ and $E^{-3}$, respectively -- we compare to the six-year Southern Sky results because that search also excluded the repeater and tested the majority of single-burst FRBs available at the time.
In the max-burst search (bottom panels), the same scale of improvement is made on the maximum flux among 28 sources at the smallest $\Delta T$.\\

\begin{table}[t]
\caption{Upper limits (90\% C.L.) on the time-integrated $E^{-2}$ power-law flux from 28 FRBs are shown in GeV$\cdot$cm$^{-2}$ for two emission timescales. This high-energy analysis sets upper limits on FRB neutrino emission on timescales similar to the radio emission (${\Delta T = 30\textrm{~ms}}$), and we compare these to limits set by ANTARES with ${\Delta T = 12\textrm{~h}}$ for twelve FRBs \citep{ANTARES:2017iky}.}
\vspace{.1in}
\centering
\begin{tabular}{ c c c }
\hline
\hline
FRB & IceCube U.L. & ANTARES U.L. \\
\hline
\hline
FRB 110220 & 0.0258 & - \\
\hline
FRB 110523 & 0.0206 & - \\
\hline
FRB 110626 & 0.112 & - \\
\hline
FRB 110703 & 0.0204 & - \\
\hline
FRB 120127 & 0.0353 & - \\
\hline
FRB 121002 & 0.175 & - \\
\hline
FRB 130626 & 0.0222 & - \\
\hline
FRB 130628 & 0.0208 & - \\
\hline
FRB 130729 & 0.0208 & - \\
\hline
FRB 131104 & 0.129 & 1.1 \\
\hline
FRB 140514 & 0.0262 & 1.9 \\
\hline
FRB 150215 & 0.0204 & 2.3 \\
\hline
FRB 150418 & 0.0390 & 1.7 \\
\hline
FRB 150610 & 0.105 & - \\
\hline
FRB 150807 & 0.134 & 1.6 \\
\hline
FRB 151206 & 0.0212 & 1.3 \\
\hline
FRB 151230 & 0.0203 & 1.6 \\
\hline
FRB 160102 & 0.0755 & 2.0 \\
\hline
FRB 160317 & 0.0689 & 1.6 \\
\hline
FRB 160410 & 0.0213 & 1.5 \\
\hline
FRB 160608 & 0.0995 & 2.1 \\
\hline
FRB 170107 & 0.0213 & 1.1 \\
\hline
FRB 170827 & 0.147 & - \\
\hline
FRB 170922 & 0.0214 & - \\
\hline
FRB 171209 & 0.124 & - \\
\hline
FRB 180301 & 0.0210 & - \\
\hline
FRB 180309 & 0.0832 & - \\
\hline
FRB 180311 & 0.139 & - \\
\hline
\end{tabular}
\label{tab:limits_perFRB}
\vspace{.1in}
\end{table}

\section{Search for Coincident MeV Neutrino Data}
\label{sec:sndaq}

IceCube is primarily designed to detect neutrinos with energies greater than 100 GeV, targeting sources with TeV neutrino emission. However, IceCube can measure a large burst of MeV neutrinos by detecting a collective rise in all photomultiplier rates on the top of the background noise. Although the increase in the counting 
rate in each light sensor is not statistically significant, the effect will be clearly seen once the rise is considered collectively over many sensors. This technique was originally developed for searches for $\mathcal{O}$(10 MeV) neutrinos from supernovae. 
IceCube utilizes a realtime data stream called the Supernova Data Acquisition (SNDAQ) system to identify collective rises in the rates of photomultipliers across the detector \citep{Abbasi:2011ss}. We use the SNDAQ data stream to search for MeV neutrinos from FRBs. 

The signals from PMTs, also called DOM hits, are counted in 2 ms bins by SNDAQ. Here, we use this data stream and search for neutrino signals from the 21 FRBs for which data were available. 
 In this search different bursts of the repeating FRB (FRB121102) are considered as individual sources. 

\subsection{Analysis method}
In order to find an excess on top of the background noise rate in the detector, a one-dimensional Gaussian likelihood is used to determine the significance of a collective deviation $(\Delta\mu)$ of the noise across the detector. 

\begin{equation}
	\mathcal{L}(\Delta\mu) = \prod_{i=1}^{N_\mathrm{DOM}} \, \frac{1}{\sqrt{2\pi}\,\langle\sigma_i\rangle} \, {\rm exp}(-\frac{(n_i-(\mu_i+\epsilon_i\,\Delta\mu))^2}{2\langle\sigma_i\rangle^2}),
\end{equation}

where $n_i$ is the per DOM$_i$ rate in a chosen time bin, $\epsilon_i$ is a DOM-specific efficiency parameter that accounts for module and depth dependent detection probabilities, and  $\mu_i$ and $\sigma_i$ are the mean and standard deviation for individual DOMs. Maximizing the log-likelihood with respect to $\Delta\mu$, one finds

\begin{equation}
	\Delta\mu = \sigma_{\Delta\mu}^2 \sum_{i=1}^{N_\mathrm{DOM}} \, \frac{\epsilon_i\,(n_i - \mu_i)}{\langle\sigma_i\rangle^2},
\end{equation}

where  
 
\begin{equation}
	\sigma_{\Delta\mu}^2 = \left(\sum_{i=1}^{N_\mathrm{DOM}} \, \frac{{\epsilon_i}^2}{\langle\sigma_i\rangle^2}\right)^{-1} .
\end{equation}

If there is an excess in the rate across the detector, its significance, $\xi$, will be given by

\begin{equation}
    \xi = \frac{\Delta\mu}{\sigma_{\Delta\mu}}.
\end{equation}

In order to search for an increase of hits during the FRB period, we bin the data collected in the SNDAQ stream into bins of 10 ms. We search in 8 different time windows from the lowest 10 ms and extending by powers of 2 up to a time window of 1280  ms. To estimate the background, we use data from a 10240 ms background-only time-window using the 8 hour runs before and after the actual FRB trigger. The background window excludes the signal time window and its size does not change as we expand the signal window.

The distribution of the significance over a course of a run in IceCube is almost a Gaussian. We use the distribution of the significances, obtained from off-time windows before and after the run that includes each FRB to obtain a threshold beyond which the significance would not arise from a random fluctuation of the background in the detector. We set a 3$\sigma$ threshold (one-sided) in the significance to claim a discovery. 

It has been shown that the rate of the hits in SNDAQ contains a contribution that is directly correlated with the seasonally changing rate of atmospheric muons traversing the detector.
In order to remove this correlation, we subtract the muon dependency via linear regression as described in \citep{Aartsen:2015cwa}. The $3\sigma$ threshold is re-evaluated according to the corrected distribution. If the significance is found greater than the threshold set, we consider that as a detection. Otherwise, we set upper limits for the absence of signal above the threshold.

\subsection{Results}

After obtaining the threshold for all time windows for the 21 FRBs considered in this analysis (see Table \ref{tab:FRBs}), we perform the likelihood analysis on the on-time window for each FRB. 

\begin{figure*}[h]
\centering
  \begin{minipage}[b]{0.49\textwidth}
  \centering
  \includegraphics[width=\textwidth]{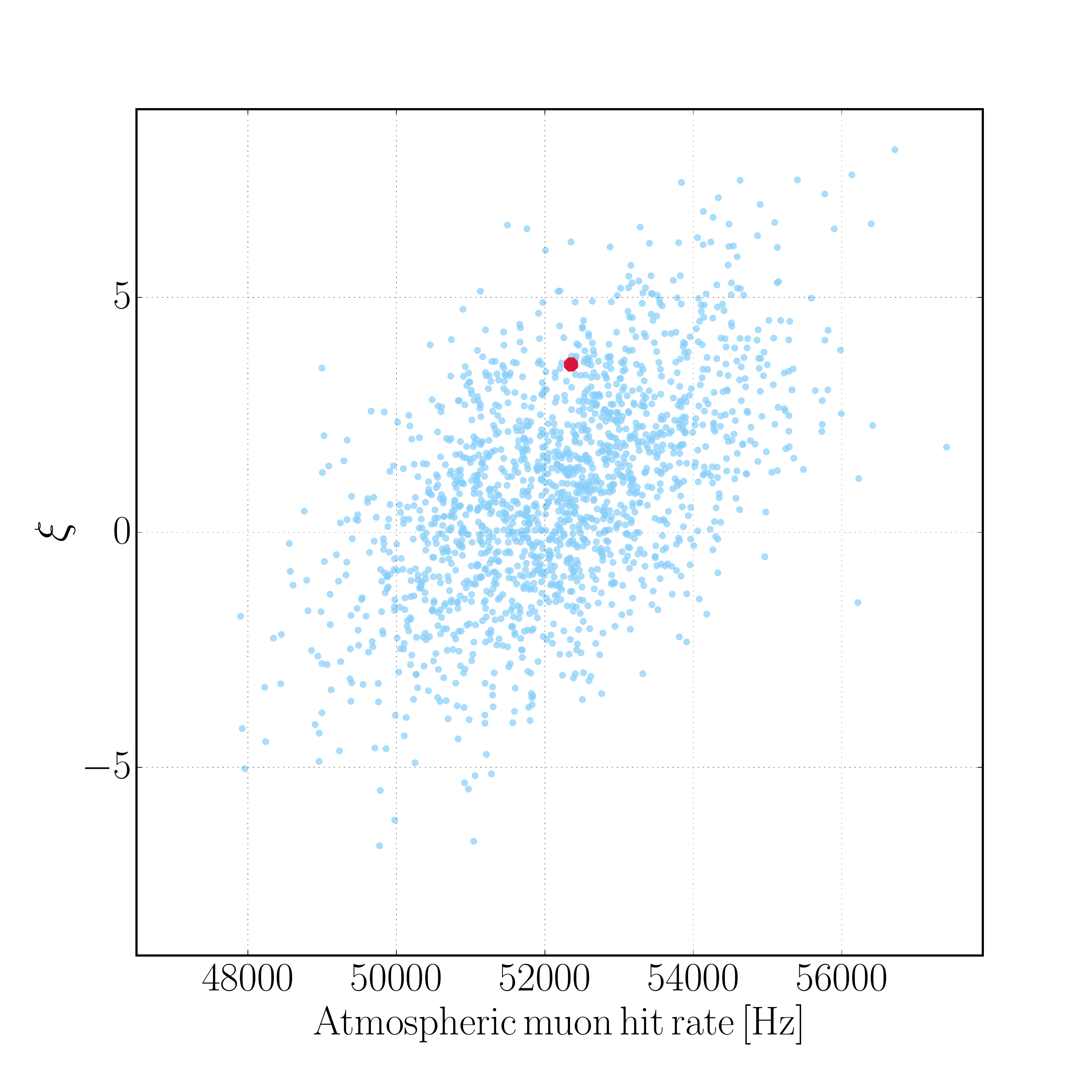}
  \end{minipage}
  \hfill
  \begin{minipage}[b]{0.49\textwidth}  
  \centering 
  \includegraphics[width=\textwidth]{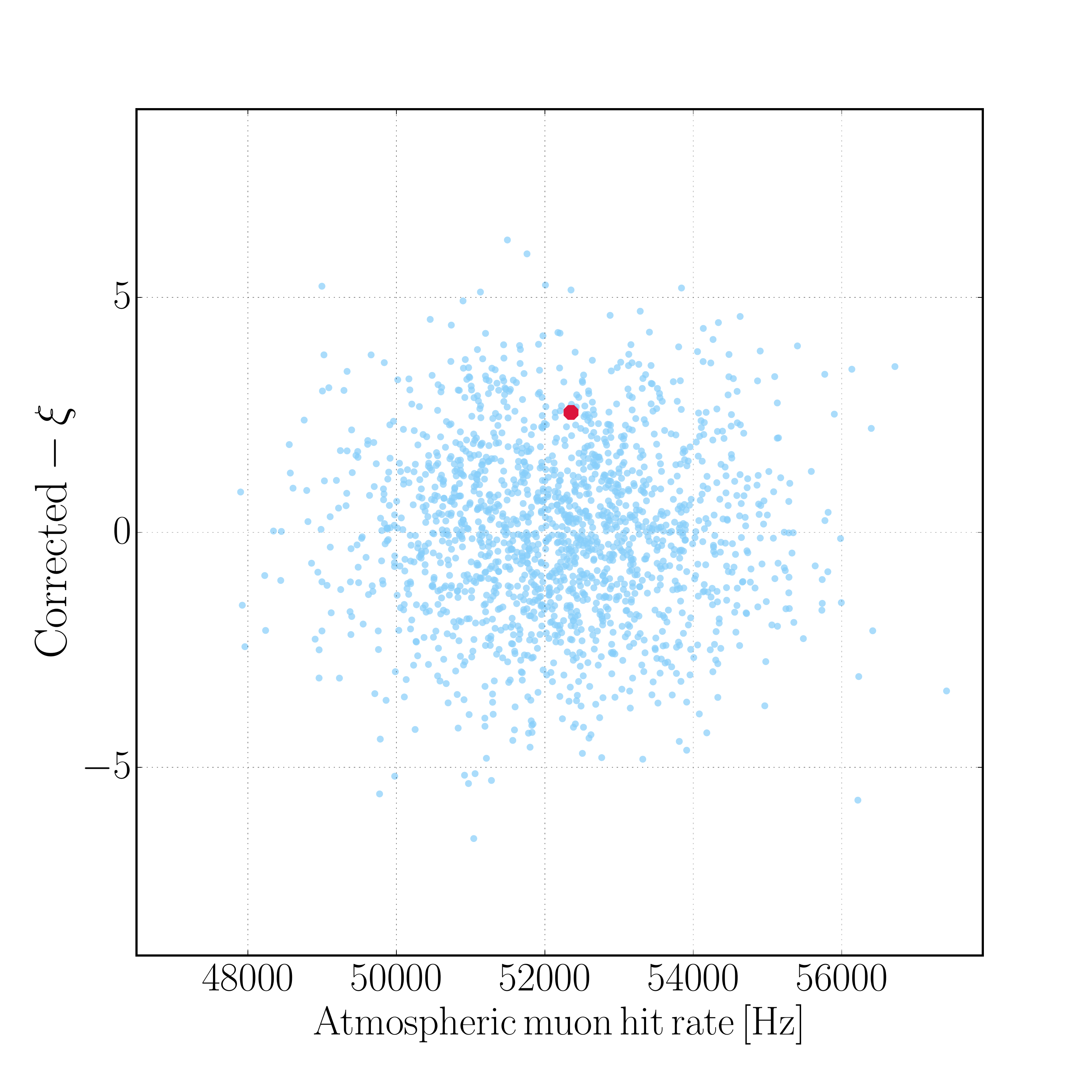}
  \end{minipage}
  \caption{Correlation of the significance ($\xi$) with atmospheric muon hit rates for the highest significance found in the sample (FRB 123202 with 1280 ms time window). The original significances (left) and corrected significances (right). The observed significance for the time of FRB is marked in red and off-time significances are marked in blue.}
  \label{fig:sig_hit}
\end{figure*}

\begin{figure}[h]
\centering
  \includegraphics[width=0.5\textwidth]{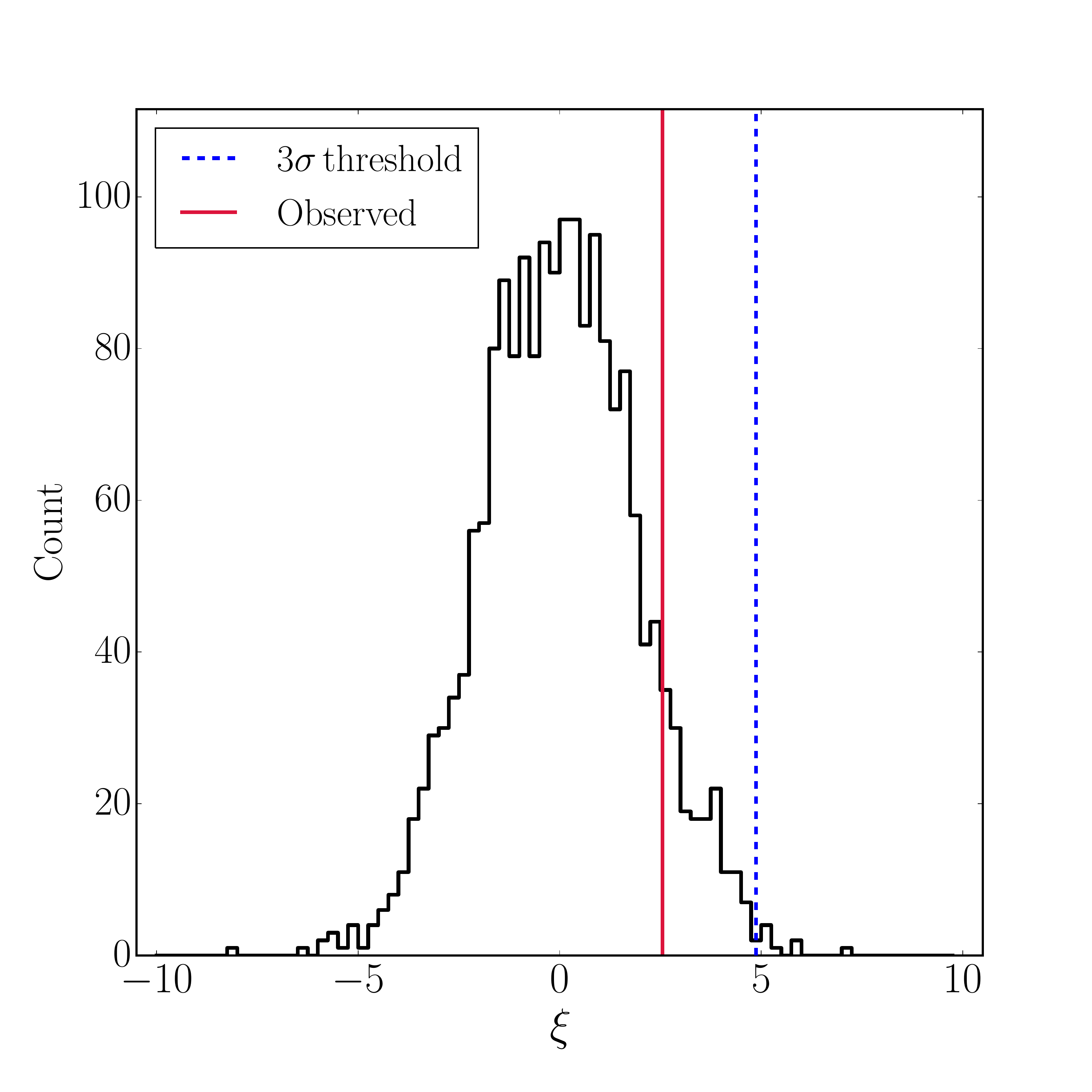}
  \label{fig:sig_distribution}
  \caption{The distribution of the significances along with the threshold (dashed) and the observed significance (red) for the most significant FRB in this search. The significance is compared to the 3$\sigma$ threshold obtained from off-time period before and after each FRB.}
\end{figure}

No significance was found above the significance threshold in the data for the time FRBs happened. The three most significant searches are presented in Table \ref{tab:SNresult}. Fig. \ref{fig:sig_hit} shows the observed significance before (left) and after (right) significance and the dependence of the significance on the atmospheric muon hit rates for the most significant FRB in the search. The distribution of the significances along with the threshold and the observed significance for the most significant search is shown in Fig. \ref{fig:sig_distribution}. Given that no results were found beyond the threshold obtained from off-time periods, we set upper limits on the flux of anti-electron neutrinos for each burst and time windows considered in this study. 

The dependency of the signal hit rate on the flux of neutrinos is described in \citep{Abbasi:2011ss}.  For the purpose of this analysis, in the absence of neutrino spectrum models for FRBs, we consider the neutrino emission from core collapse supernova as a fiducial model to obtain the upper limit on MeV neutrino emission. Here, the normalization was chosen such that it corresponds to a model describing the neutrino flux with average neutrino energy $E_\nu=15.6$ MeV and pinching parameter $\alpha=3$ \citep{Totani:1997vj}, yielding  $<E_\nu^3> = 7118\, \rm MeV^3$.
 In order to find the upper limit on the neutrino flux at MeV energies, we evaluate the required time-integrated flux of anti-electron neutrinos that would produce an enhancement in the signal rate in the detector corresponding to 90\% one-sided confidence level for the Gaussian distribution of the significance in the off-time runs. The 90\% one-sided confidence level for a Gaussian in $\xi$ is given  by
\begin{eqnarray}
\xi^{90} &=&  \xi_\mathrm{th} + z_{90}\cdot \sigma_\xi ,
\end{eqnarray}

where $\xi_\mathrm{th}$ is the significance threshold, $\sigma_\xi$ is the width of the significance distribution, and $z_{90}\approx 1.282$ is the fraction of the gaussian width (1 sigma) representing the 90\% confidence level interval. $\xi^{90}$ is the significance for which only in 10\% of random draws one finds $\xi < \xi_\mathrm{cut}$. The corresponding additional hits per DOM will be 

\begin{eqnarray}
\Delta\mu^{90} & = & \sigma_{\Delta\mu}\left(\xi_\mathrm{th} + z_{90}\cdot \sigma_\xi\right).
\end{eqnarray}

\begin{figure}
\centering
  \includegraphics[width=0.77\textwidth]{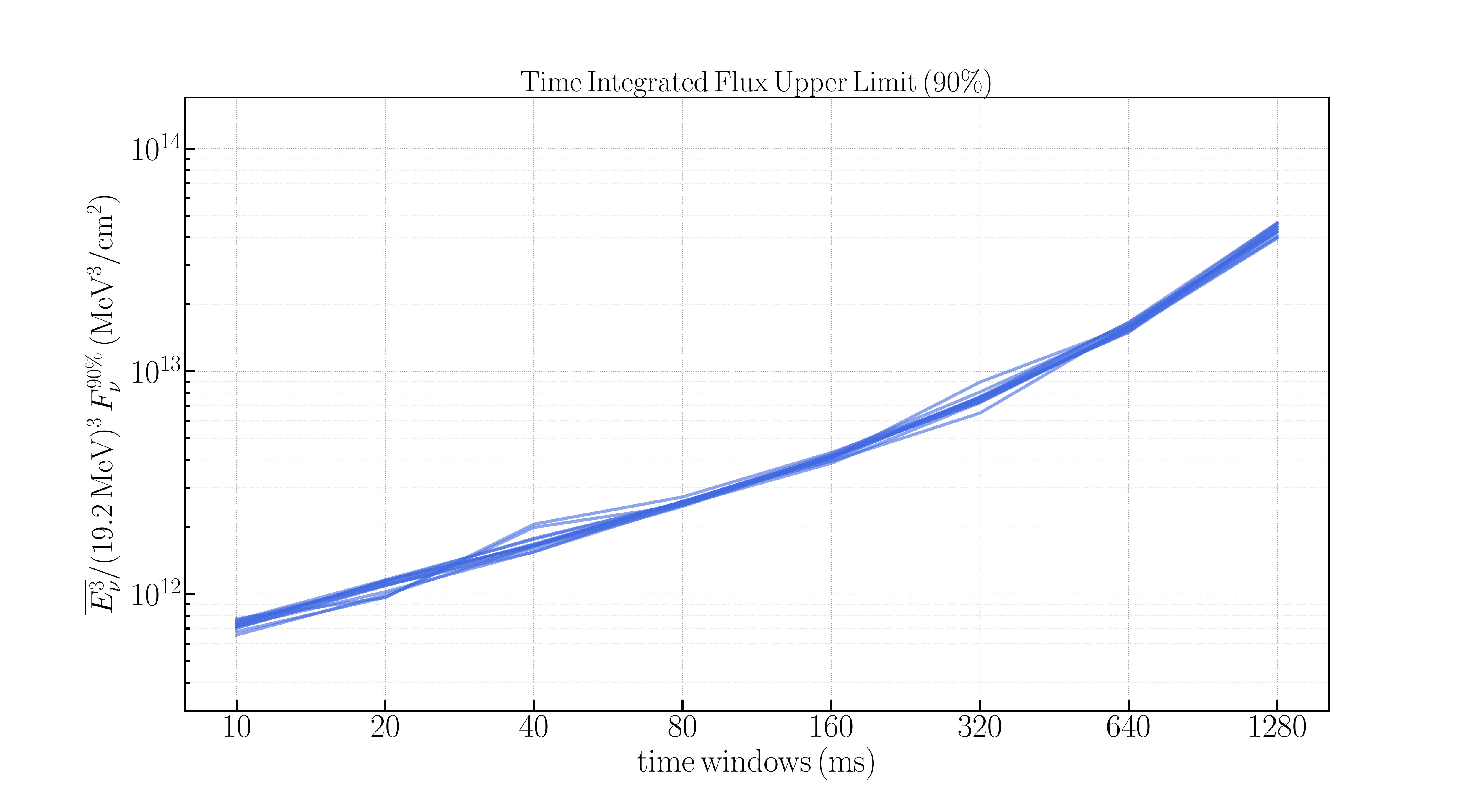}
  \label{fig:upperlimit}
  \caption{{90\% C.L. on the time-integrated flux of anti-electron neutrinos from 21 FRBs considered in the analysis, assuming the fiducial model for neutrino emission with a mean neutrino energies of 15.6 MeV.}}
\end{figure}

The total increase in the signal hit rates in the detector is obtained by adding the signal of all DOMs. 
The upper limit on the time-integrated flux for FRBs considered in this analysis are shown in Figure~\ref{fig:upperlimit}.

\begin{table}[t]
\caption{SNDAQ search results for the most significant bursts in the analysis.}
\vspace{.1in}
\centering
\begin{tabular}{ c c c c c}
\hline
\hline
FRB &  TW (ms)	& Significance (before muon correction) & Significance (after muon correction)\\
\hline
121102b1 & 	640 & 	2.97 &	0.88\\
\hline
121102b1 &	1280 &	3.03 &	0.85\\
\hline
131104 &  1280 & 3.57	& 2.55\\
\hline
\end{tabular}
\label{tab:SNresult}
\vspace{.3in}
\end{table}

\section{Conclusion and Discussion}
\label{sec:conclusion}

In two searches for neutrino emission from FRBs -- one for track-like events from muon neutrinos above 100 GeV and the other for MeV neutrino events -- no significant association has been found.

We set upper limits (90\% confidence level) on the time-integrated neutrino flux from FRBs that are the most constraining to date. For a power-law spectrum $E^{-2}$ ($E^{-3}$), the limit set for GeV-TeV neutrinos is ${E^2 F < 2\times10^{-3}\mathrm{~GeV~cm}^{-2}}$ (${E^2 F < 2\times10^{-4}\mathrm{~GeV~cm}^{-2}}$ at 100 TeV) per burst for emission timescales less than 10 seconds. We also set the first upper limits on MeV neutrino emission from FRBs: ${<E^3> F <8 \times 10^{11}\mathrm{~MeV^3~cm}^{-2}}$ at an emission timescale of 10 ms (Figure~\ref{fig:upperlimit}).

In the tracks search, IceCube is more sensitive in the Northern sky than in the Southern, where most FRB sources have been detected so far. Additionally, stacking sensitivity scales roughly with the number of sources for which detection of a spatially coincident background event within the time window is unlikely. Therefore, as more FRBs are detected, especially in the northern sky by observatories like CHIME \citep{Amiri:2018qsq,chimefrb}, the stacking analysis sensitivity can improve by orders of magnitude. With fields of view significantly larger than for single-dish telescopes, new radio interferometers may detect several FRBs per day in the coming years \citep{Amiri:2018qsq, Newburgh:2016mwi}, accelerating this improvement.

The MeV neutrino search was made possible by the ability of IceCube to identify burst of MeV neutrinos on top of the background noise in the detector. While this capability has been primarily incorporated for obtaining early alerts on supernova explosions, it offers a unique opportunity for temporal study of low-energy neutrino emission from transients. With this opportunity, we have placed the first ever limits on neutrino signals at MeV energies from FRBs. Prsopects for observation of an excess of MeV neutrinos in IceCube depends on the distance to the source. While IceCube is highly sensitive for identification of MeV signals from Galactic distances, there is growing evidence that most FRBs are extragalactic now that the redshift has been measured for two sources\citep{Bannistereaaw5903,ravi2019}. It is worth mentioning that in addition, on request, all untriggered DOM hits in the detector may be stored for any particular time period for several days \citep{Aartsen:2016nxy}. IceCube has used this channel to search for neutrino emission from gravitational wave sources. For more details, see \citep{ANTARES:2017iky}

\section*{acknowledgements} The IceCube collaboration acknowledges the significant contributions to this manuscript from Sam Fahey, Ali Kheirandish, and Justin Vandenbroucke. The authors gratefully acknowledge the support from the following agencies and institutions: USA {\textendash} U.S. National Science Foundation-Office of Polar Programs, U.S. National Science Foundation-Physics Division, Wisconsin Alumni Research Foundation, Center for High Throughput Computing (CHTC) at the University of Wisconsin-Madison, Open Science Grid (OSG), Extreme Science and Engineering Discovery Environment (XSEDE), U.S. Department of Energy-National Energy Research Scientific Computing Center, Particle astrophysics research computing center at the University of Maryland, Institute for Cyber-Enabled Research at Michigan State University, and Astroparticle physics computational facility at Marquette University; Belgium {\textendash} Funds for Scientific Research (FRS-FNRS and FWO), FWO Odysseus and Big Science programmes, and Belgian Federal Science Policy Office (Belspo); Germany {\textendash} Bundesministerium f{\"u}r Bildung und Forschung (BMBF), Deutsche Forschungsgemeinschaft (DFG), Helmholtz Alliance for Astroparticle Physics (HAP), Initiative and Networking Fund of the Helmholtz Association, Deutsches Elektronen Synchrotron (DESY), and High Performance Computing cluster of the RWTH Aachen; Sweden {\textendash} Swedish Research Council, Swedish Polar Research Secretariat, Swedish National Infrastructure for Computing (SNIC), and Knut and Alice Wallenberg Foundation; Australia {\textendash} Australian Research Council; Canada {\textendash} Natural Sciences and Engineering Research Council of Canada, Calcul Qu{\'e}bec, Compute Ontario, Canada Foundation for Innovation, WestGrid, and Compute Canada; Denmark {\textendash} Villum Fonden, Danish National Research Foundation (DNRF), Carlsberg Foundation; New Zealand {\textendash} Marsden Fund; Japan {\textendash} Japan Society for Promotion of Science (JSPS) and Institute for Global Prominent Research (IGPR) of Chiba University; Korea {\textendash} National Research Foundation of Korea (NRF); Switzerland {\textendash} Swiss National Science Foundation (SNSF); United Kingdom {\textendash} Department of Physics, University of Oxford.
\bibliographystyle{yahapj}
\bibliography{references}

\begin{table}[t]
\caption{39 FRBs from 29 unique directions (repeated bursts from FRB 121102 are labelled with an additional "b0", "b1", etc.) are included in the analyses presented here. Checkmarks ($\checkmark$) indicate that an FRB is analyzed in the MeV ("SNDAQ") and/or GeV-TeV ("Tracks") stream. The latter included all non-repeating FRBs. Additional burst characteristics were taken from www.frbcat.org \citep{Petroff:2016tcr}: arrival time and duration corrected for signal dispersion, right ascension, and declination (J2000) rounded to the nearest arcminute.}
\vspace{.1in}
\centering
\begin{tabular}{ c c c c c c c }
\hline
\hline
FRB & SNDAQ & Tracks & Time (UTC) & Duration (ms) & RA & DEC \\
\hline
\hline
FRB 110220 & \checkmark & \checkmark & 2011-02-20 01:55:48.957 & 5.6 & 22h 34$^\prime$ & -12$^\circ$ 24$^\prime$ \\
\hline
FRB 110523 & \checkmark & \checkmark & 2011-05-23 15:06:19.738 & 1.73 & 21h 45$^\prime$ & -00$^\circ$ 12$^\prime$ \\
\hline
FRB 110626 & \checkmark & \checkmark & 2011-06-26 21:33:17.474 & $<1.4$ & 21h 03$^\prime$ & -44$^\circ$ 44$^\prime$ \\
\hline
FRB 110703 & & \checkmark & 2011-07-03 18:59:40.591 & $<4.3$ & 23h 30$^\prime$ & -02$^\circ$ 52$^\prime$ \\
\hline
FRB 120127 & \checkmark & \checkmark & 2012-01-27 08:11:21.723 & $<1.1$ & 23h 15$^\prime$ & -18$^\circ$ 25$^\prime$ \\
\hline
FRB 121002 & \checkmark & \checkmark & 2012-10-02 13:09:18.402 & 2.1; 3.7 & 18h 14$^\prime$ & -85$^\circ$ 11$^\prime$ \\
\hline
FRB 121102 b0& \checkmark & & 2012-11-02 06:47:17.117 & 3.3 & 05h 32$^\prime$ & +33$^\circ$ 05$^\prime$ \\
\hline
FRB 130626 & \checkmark & \checkmark & 2013-06-26 14:56:00.06 & $<0.12$ & 16h 27$^\prime$ & -07$^\circ$ 27$^\prime$ \\
\hline
FRB 130628 & \checkmark & \checkmark & 2013-06-28 03:58:00.02 & $<0.05$ & 09h 03$^\prime$ & +03$^\circ$ 26$^\prime$ \\
\hline
FRB 130729 & \checkmark & \checkmark & 2013-07-29 09:01:52.64 & $<4$ & 13h 41$^\prime$ & -05$^\circ$ 59$^\prime$ \\
\hline
FRB 131104 & \checkmark & \checkmark & 2013-11-04 18:04:01.2 & $<0.64$ & 06h 44$^\prime$ & -51$^\circ$ 17$^\prime$ \\
\hline
FRB 140514 & & \checkmark & 2014-05-14 17:14:11.06 & 2.8 & 22h 34$^\prime$ & -12$^\circ$ 18$^\prime$ \\
\hline
FRB 150215 & & \checkmark & 2015-02-15 20:41:41.714 & 2.88 & 18h 17$^\prime$ & -4$^\circ$ 54$^\prime$ \\
\hline
FRB 150418 & \checkmark & \checkmark & 2015-04-18 04:29:05.370 & 0.8 & 07h 16$^\prime$ & -19$^\circ$ 00$^\prime$ \\
\hline
FRB 121102 b1& \checkmark & & 2015-05-17 17:42:08.712 & 3.8 & 05h 32$^\prime$ & +33$^\circ$ 05$^\prime$ \\
\hline
FRB 121102 b2& \checkmark & & 2015-05-17 17:51:40.921 & 3.3 & 05h 32$^\prime$ & +33$^\circ$ 05$^\prime$ \\
\hline
FRB 121102 b3& \checkmark & & 2015-06-02 16:38:07.575 & 4.6 & 05h 32$^\prime$ & +33$^\circ$ 05$^\prime$ \\
\hline
FRB 121102 b4& \checkmark & & 2015-06-02 16:47:36.484 & 8.7 & 05h 32$^\prime$ & +33$^\circ$ 05$^\prime$ \\
\hline
FRB 121102 b5& \checkmark & & 2015-06-02 17:49:18.627 & 2.8 & 05h 32$^\prime$ & +33$^\circ$ 05$^\prime$ \\
\hline
FRB 121102 b6& \checkmark & & 2015-06-02 17:49:41.319 & 6.1 & 05h 32$^\prime$ & +33$^\circ$ 05$^\prime$ \\
\hline
FRB 121102 b7& \checkmark & & 2015-06-02 17:50:39.298 & 6.6 & 05h 32$^\prime$ & +33$^\circ$ 05$^\prime$ \\
\hline
FRB 121102 b8& \checkmark & & 2015-06-02 17:53:45.528 & 6.0 & 05h 32$^\prime$ & +33$^\circ$ 05$^\prime$ \\
\hline
FRB 121102 b9& \checkmark & & 2015-06-02 17:56:34.787 & 8.0 & 05h 32$^\prime$ & +33$^\circ$ 05$^\prime$ \\
\hline
FRB 121102 b10& \checkmark & & 2015-06-02 17:57:32.020 & 3.1 & 05h 32$^\prime$ & +33$^\circ$ 05$^\prime$ \\
\hline
FRB 150610 & & \checkmark & 2015-06-10 05:26:59.396 & 2.00 & 10h 44$^\prime$ & -40$^\circ$ 05$^\prime$ \\
\hline
FRB 150807 & & \checkmark & 2015-08-07 17:53:55.83 & 0.35 & 22h 43$^\prime$ & -55$^\circ$ 05$^\prime$ \\
\hline
FRB 151206 & & \checkmark & 2015-12-06 06:17:52.778 & 3.00 & 19h 21$^\prime$ & -04$^\circ$ 08$^\prime$ \\
\hline
FRB 151230 & & \checkmark & 2015-12-30 16:15:46.525 & 4.40 & 09h 40$^\prime$ & -03$^\circ$ 27$^\prime$ \\
\hline
FRB 160102 & & \checkmark & 2016-01-02 08:28:39.374 & 3.40 & 22h 39$^\prime$ & -30$^\circ$ 11$^\prime$ \\
\hline
FRB 160317 & & \checkmark & 2016-03-17 09:00:36.53 & 21.00 & 07h 54$^\prime$ & -29$^\circ$ 37$^\prime$ \\
\hline
FRB 160410 & & \checkmark & 2016-04-10 08:33:39.68 & 4.00 & 08h 41$^\prime$ & +06$^\circ$ 05$^\prime$ \\
\hline
FRB 160608 & & \checkmark & 2016-06-08 03:53:01.088 & 9.00 & 07h 37$^\prime$ & -40$^\circ$ 48$^\prime$ \\
\hline
FRB 170107 & & \checkmark & 2017-01-07 20:05:45.139 & 2.60 & 11h 23$^\prime$ & -05$^\circ$ 00$^\prime$ \\
\hline
FRB 170827 & & \checkmark & 2017-08-27 16:20:18 & 0.40 & 00h 49$^\prime$ & -65$^\circ$ 33$^\prime$ \\
\hline
FRB 170922 & & \checkmark & 2017-09-22 11:22:23.40 & 26.00 & 21h 30$^\prime$ & -08$^\circ$ 00$^\prime$ \\
\hline
FRB 171209 & & \checkmark & 2017-12-09 20:34:23.50 & 2.50 & 15h 50$^\prime$ & -46$^\circ$ 10$^\prime$ \\
\hline
FRB 180301 & & \checkmark & 2018-03-01 07:34:19.76 & 3.00 & 06h 13$^\prime$ & +04$^\circ$ 34$^\prime$ \\
\hline
FRB 180309 & & \checkmark & 2018-03-09 02:49:32.99 & 0.58 & 21h 25$^\prime$ & -33$^\circ$ 59$^\prime$ \\
\hline
FRB 180311 & & \checkmark & 2018-03-11 04:11:54.80 & 12.00 & 21h 32$^\prime$ & -57$^\circ$ 44$^\prime$ \\
\hline
\end{tabular}
\label{tab:FRBs}
\vspace{.3in}
\end{table}
\end{document}